\documentclass[preprint,12pt]{elsarticle}

\usepackage[margin=2.5cm]{geometry} 
\usepackage{amssymb}
\usepackage{amsmath}
\usepackage{xcolor}

\usepackage{soul}
\usepackage{ulem} 
\usepackage{tabularx}
\usepackage{graphicx}
\usepackage{dcolumn}
\usepackage{bm}
\usepackage{natbib}
\usepackage[hidelinks]{hyperref}

\begin{document}

\begin{frontmatter}

\title{Phonon modal analysis of thermal transport in ThO$_2$ with point defects using equilibrium molecular dynamics}
 
\author[1]{Beihan Chen}
\author[2]{Linu Malakkal}
\author[3]{Marat Khafizov}
\author[4]{David H. Hurley}
\author[1]{Miaomiao Jin\corref{cor1}}
\ead{mmjin@psu.edu}
\cortext[cor1]{Corresponding author}

\address[1]{Department of Nuclear Engineering, The Pennsylvania State University, University Park, PA 16802, USA}
\address[2]{Computational Mechanics and Materials Department, Idaho National Laboratory, Idaho Falls, ID 83415, USA}
\address[3]{Department of Mechanical and Aerospace Engineering, The Ohio State University, Columbus, OH 43210, USA}
\address[4]{Department of Materials Science and Engineering, Idaho National Laboratory, Idaho Falls, ID 83415, USA}

\begin{abstract}
Defects can significantly degrade the thermal conductivity of ThO$_2$, an advanced nuclear fuel material as well as a surrogate for other fluorite-structured materials. We investigate how point defects in ThO$_2$ impact phonon mode-resolved thermal transport. By incorporating phonon modes from lattice dynamics, we decompose the trajectory and heat flux to phonon normal mode space and extract key phonon properties, including phonon relaxation times and their contributions to thermal conductivity. We implement two methods. The first method is based on the Green Kubo formalism to resolve the contribution of each phonon mode to thermal conductivity. The second resolves the lifetime of individual phonon modes and the thermal conductivity is calculated using the Boltzmann transport equation within relaxation time approximation. Notably, a lower contribution of acoustic modes is revealed compared to perturbative approaches considering only three-phonon scattering processes. The effects of four types of point defects are evaluated. The strongest impact on a reduction in thermal conductivity is from Th interstitials, followed by Th vacancies. O interstitials/vacancies have a similar impact, albeit smaller than defects on the thorium sublattice. These observations are consistent with previous studies.

\end{abstract}


\end{frontmatter}


\section{Introduction}

Thorium dioxide (ThO$_2$) shows great potential for use in nuclear fuels due to its enhanced thermo-physical properties, reduced waste by volume compared to conventional uranium dioxide (UO$_2$) fuel \cite{Lombardi2008}, and increased resistance against potential nuclear weapons proliferation \cite{Ashley2012}. Thermal transport in nuclear fuel plays a pivotal role in determining the efficiency and safety of nuclear reactors. Therefore, establishing a fundamental and comprehensive understanding of thermal transport in ThO$_2$ and related fluorite oxides in reactor conditions is imperative for nuclear systems \cite{Hurley2022,Dennett2021}. It is well known that irradiation-induced defects, including points defects, fission products, and defect clusters, can significantly degrade the thermal conductivity in fluorite oxides, whose conductivity is governed by phonons, due to enhanced phonon-defect scattering \cite{Jin2022,Deskins2022}. Given the limitations in experimental sample preparation and handling, physics-based modeling has played a substantial role in predicting thermal conductivity across various temperature ranges and under the influence of different types of defects \cite{Dennett2021}.

Previous studies have presented a rich analysis of thermal conductivity in pristine ThO$_2$, encompassing both experimental and computational investigations \cite{Bakker1997,Malakkal2019,Park2018bulk,COOPER201529,HUA2023}. Bakker et al. \cite{Bakker1997} collected experimental data of ThO$_2$ thermal conductivity. Additionally, computational simulations using equilibrium molecular dynamics (EMD) \cite{Malakkal2019,MA2015476} and non-equilibrium molecular dynamics (NEMD) \cite{fmuller1997,Park2018bulk} methods are employed to predict the value of ThO$_2$ thermal conductivity. In both experimental and computational works, the thermal conductivity of ThO$_2$ at different temperatures has been investigated. A broad range of values has been reported. For example, Malakkal et al. \cite{Malakkal2019} reported 16.89 W/mK at 300 K and 7.55 W/mK at 670 K from the experiment, while Jin et al. \cite{Jin2021} reported 20.84 W/mK at 300 K and 8.86 W/mK at 700 K from NEMD simulation. 

With the existence of defects, phonon-defect scattering can be described as a perturbation to harmonic lattice dynamics \cite{Gurunathan2020}. Up to now, the widely accepted description of phonon-phonon and phonon-defect scattering is based on the seminal work by Klemens \cite{Klemens1955, klemens1958thermal} and Callaway \cite{Callaway1959}. Thermal transport of ThO$_2$ with point defects has been studied using NEMD simulation in previous works \cite{Park2018defect,Jin2022}. The collective findings from these NEMD studies consistently indicate that point defects can significantly impact the thermal transport properties of ThO$_2$. For example, it was reported that at 300 K, 0.1\% of oxygen vacancy could cause a 20\% degradation in the thermal conductivity at 300 K \cite{Park2018defect}. The NEMD method can obtain thermal conductivity directly and simulate very large systems but lacks a comprehensive phonon mode-resolved exploration of the underlying mechanisms for thermal conductivity degradation. Recently, the phonon-defect scattering rate was estimated using Green’s function T-matrix method \cite{Mingo2010,Kundu2011} based on interatomic force constants (IFCs) \cite{gonze1994interatomic} from first principles to evaluate the impact of point defects on the thermal conductivity of ThO$_2$ \cite{Malakkal2024}. The T-matrix method considers only three-phonon scattering processes without higher-order anharmonic effects, and only oxygen and thorium vacancies ($\mathrm{V_{O}}$ and $\mathrm{V_{Th}}$) were investigated, as the T-matrix method to capture the characteristic of interstitial defects needs further development \cite{Jiang2022}. In addition, Deskins et al. \cite{Deskins2021} solved the Boltzmann transport equation (BTE) to study the impact of point defects on thermal conductivity in ThO$_2$, where the Klemens' model \cite{Klemens1955} was used to evaluate the phonon scattering cross-section \cite{Deskins2022}.

Here, we conduct phonon modal analysis on the thermal transport of pristine ThO$_2$ systems and ThO$_2$ systems containing point defects, including oxygen vacancies ($\mathrm{V_{O}}$), oxygen interstitials ($\mathrm{I_O}$), thorium vacancies ($\mathrm{V_{Th}}$), and thorium interstitials ($\mathrm{I_{Th}}$). The Green-Kubo modal analysis (GKMA) \cite{Lv2016} and the relaxation time approximation with normal mode analysis (RTA-NMA) \cite{Feng2015} methods are employed based on lattice dynamics (LD) and EMD simulation. The GKMA method has been utilized in various materials to directly compute phonon modal contributions to thermal conductivity \cite{Seyf2019}. Meanwhile, the RTA-NMA method, despite its underlying simplified assumptions on phonon transport properties such as phonon relaxation time, finds extensive application in the fields of ceramics and semiconductors for computing thermal conductivity \cite{Henry2008,Shao2016}. Such application of these two methods to study thermal transport in ThO$_2$ provides complementary information on the phonon transport in the presence of defects. To obtain detailed and reliable calculation results, the cost of computation and storage is high; hence, the size of the atomic configurations used in both the GKMA and RTA-NMA methods is limited to calculate phonon mode-resolved thermal conductivity. Detailed phonon modal analysis for defect-bearing ThO$_2$ is reported in this study, the obtained results are subject to comparative analysis under varying temperature conditions and defects at different concentrations.

\section{Methods}
Two methods that combine EMD and LD are employed to investigate the thermal transport of perfect and point-defect-bearing ThO$_2$. LD simulation is used to determine eigenvectors and other properties of phonons, while EMD simulation is used to determine phonon modal thermal conductivity.

\subsection{Green-Kubo modal analysis (GKMA) }

Green-Kubo (GK) formulation has been frequently used to study transport phenomena, based on the fluctuation-dissipation theorem \cite{Green1954,Kubo1957}. It has been used to quantify thermal conductivity in ThO$_2$ \cite{Malakkal2019}. Specifically, thermal conductivity $\kappa$ is calculated by integrating the instantaneous heat current ($Q(t)$) auto-correlation function (HCACF) \cite{Kang2017},

\begin{equation}\label{eq_gk}
    \kappa = \frac{1}{k_{B}T^{2}V} \int \langle Q(t)Q(0) \rangle dt
\end{equation}
Within the GK theory, we employ a phonon modal analysis method introduced by Lv and Henry \cite{Lv2016}, which incorporates normal mode coordinates ($X_n(t)$ in Eq. (\ref{X_n})) and velocities ($\dot{X}_n(t)$ in Eq. (\ref{Xdot_n})) into the instantaneous heat current, i.e.,
\begin{equation}\label{eq_gkma_heatflux}
Q(t)=\sum_{n=1}^{3N} Q_n(t)
\end{equation}
\begin{equation}
    Q_n(t)=\sum_{i=1}^{N} \{ E_i [ \frac{1}{m_i^{1/2}} \mathbf{e}_{i,n} \dot{X}_n(t)]+\sum_{j=1}^{N} [ - \nabla_{\mathbf{r}_i} \mathbf{\Phi}_j \cdot ( \frac{1}{m_i^{1/2}} \mathbf{e}_{i,n} \dot{X}_n(t)) ]\mathbf{r}_{ij}\}
\end{equation}
\begin{equation}\label{X_n}
X_n(t)=\sum_{j=1}^{N} \sqrt{m_j} \mathbf{e}_{j,n}^* \cdot \mathbf{u}_j(t)
\end{equation}
\begin{equation}\label{Xdot_n}
\dot{X}_n(t)=\sum_{j=1}^{N}\sqrt{m_j}\mathbf{e}_{j,n}^*\cdot\dot{\mathbf{u}}_j(t)
\end{equation}
where $i$ and $j$ is the atom index, $m$ is the mass, $\mathbf{u}(t)$ is the atomic displacement (obtained from EMD), $\mathbf{e}_{j,n}$ is the component of phonon mode $n$'s eigenvector for atom $j$ (obtained from LD), and * denotes the complex conjugate. $Q_n(t)$ is phonon modal heat flux, $E_i$ is the total energy (sum of potential and kinetic energy) of atom $i$, $\mathbf{\Phi}_j$ is the potential energy of atom $j$, and $\mathbf{r}_{ij}=\mathbf{r}_{j}-\mathbf{r}_{i}$ (the distance vector between atom $i$ and atom $j$). Plugging Eq. (\ref{eq_gkma_heatflux}) into Eq. (\ref{eq_gk}), we have

\begin{equation}\label{eq_gkma_1d}
\mathbf{\kappa}(n) = \frac{1}{k_{B}T^{2}V}  \int \langle Q_{n}(t)Q(0) \rangle dt = \frac{1}{k_{B}T^{2}V} \sum_{n^{'}=1}^{3N} \int \langle Q_{n}(t)Q_{n^{'}}(0) \rangle dt
\end{equation}
\begin{equation}\label{eq_gkma_2d}
\mathbf{\kappa} = \sum_{n=1}^{3N} \mathbf{\kappa}(n) = \frac{1}{k_{B}T^{2}V} \sum_{n=1}^{3N} \sum_{n^{'}=1}^{3N}  \int \langle Q_{n}(t)Q_{n^{'}}(0) \rangle dt 
\end{equation}
With this, each phonon mode’s contribution to the total thermal conductivity can be obtained. Furthermore, Eq. (\ref{eq_gkma_2d}) enables examination of how the cross-correlation between pairs of modes contributes to thermal conductivity.

\subsection{Relaxation-time approximation with normal mode analysis (RTA-NMA)}

Within this method, phonon relaxation times are extracted by NMA on the EMD simulation trajectory. This method has been demonstrated in previous studies, e.g., by McGaughey et al. \cite{McGaughey2004} and Turney et al. \cite{Turney2009}. The core concept of this method is to decompose the atomic trajectories obtained from EMD to phonon normal modes obtained from LD, and then the relaxation times are computed by analyzing the temporal decay of the auto-correlation function associated with the total energy of each phonon mode. Then, thermal conductivity is evaluated based on BTE under the relaxation time approximation \cite{Ladd1986,srivastava2022physics},  
\begin{equation}\label{eq_nma_kappa}
    \mathbf{\kappa} = \sum_{n=1}^{3N} c_{n}\mathbf{v}_{g,n}^{2}\tau_n
\end{equation}
where $\kappa$ is thermal conductivity, $N$ represents the number of atoms, and $n$ (ranging from 1 to $3N$) is the index of each phonon mode, according to the LD calculation, the first N modes are acoustic, the remaining 2N modes are optical \cite{Jin2021}. $\mathbf{v}_{g,n}$, $\tau_n$, and $c_{n}$ are the group velocity (obtained from LD), relaxation time, and volumetric specific heat associated with the phonon mode $n$, respectively. Specifically, $c_{n}$ is written as \cite{PORTER199753}, 

\begin{equation}\label{eq_nma_spheat}
    c_n = \frac{k_{B}}{V} \frac{x^{2}e^{x}}{(e^{x}-1)^2},  x = \frac{ \hbar \omega_n }{ k_{B} T}
\end{equation}
where $k_B$ is the Boltzmann constant, $T$ is temperature, $V$ is system volume, and $\omega_n$ is phonon frequency of mode $n$ (obtained from LD). A critical step for determining thermal conductivity is to calculate $\tau_n$, which is based on the auto-correlation function of the total energy of phonon mode $n$ ($E_n$). $E_n$ can be calculated from the time-dependent normal mode coordinate of phonon mode $n$ ($A_{n}(t)$) and its derivative to time $\dot{A}_{n}(t)$, i.e., 

\begin{equation}\label{eq_nma_Em}
E_n(t) = \omega_n^2 \frac{ A_n(t)^*A_n(t)}{2} + \frac{\dot{A}_n(t)^*\dot{A}_n(t)}{2}
\end{equation}
where $A_n(t)$ can be expressed as \cite{Henry2008},

\begin{equation}\label{eq_nma_Am}
A_{n}(t) = \sum_{j} \sqrt{\frac{m_j}{N}} \mathrm{exp}(i\mathbf{k}\cdot\mathbf{r_j(0)}) \mathbf{e}_{j,n}^* \cdot \mathbf{u}_{j}(t) 
\end{equation}
Then, the relaxation time of phonon mode $n$ can be obtained by exponential fitting of normalized auto-correlation of its total energy $E_n(t)$:

\begin{equation}\label{eq_nma_ac}
\mathrm{exp}(-\frac{t}{\tau_n}) = \frac{ \langle E_{n}(t)E_{n}(0) \rangle } { \langle E_{n}(0)E_{n}(0) \rangle} .
\end{equation}

The cumulative thermal conductivity $\kappa_{C}(\omega)$ is determined via aggregated contribution to thermal conductivity from phonon modes within a frequency range starting from 0 THz up to a specified frequency $\omega$,
\begin{equation}\label{eq_nma_cumulative}
\kappa_{C}(\omega) = \sum_{0<\omega_{n}\leqslant\omega}\kappa_{n}(\omega_n),~ \text{where }   \kappa_{n}(\omega_n)=c_{n}(\omega_n)\mathbf{v}_{g,n}^{2}\tau_n(\omega_n)
\end{equation}

\subsection{Simulation setups}

For the EMD simulations of ThO$_2$, we utilize the Large-scale Atomic/Molecular Massively Parallel Simulator (LAMMPS) \cite{LAMMPS} package, combined with an empirical interatomic potential (EIP) referred to as the Cooper-Rushton-Grimes (CRG) potential from Cooper et al. \cite{Cooper2014}. This potential has been demonstrated well to reproduce thermophysical and defect properties in ThO$_2$ \cite{Cooper2014,cooper2015modellingthermal,cooper2015thermophysical}. The General Utility Lattice Program (GULP) \cite{Gale2003} is used for LD calculations to obtain the phonon eigenvectors $\mathbf{e}_{j,n}$, frequencies $\omega_{n}$ and group velocities $\mathbf{v}_{g,n}$ for the perfect lattice at the ground state. A ThO$_2$ supercell structure containing 1500 atoms (5$\times$5$\times$5 conventional unit cells) is used throughout this work. Such choice of supercell size is based on a converging test on the thermal conductivity using supercell sizes ranging from 3$\times$3$\times$3 to 6$\times$6$\times$6 conventional unit cells (see details in Supplementary Materials (SM), Figure S1) and the trade-off between phonon details and computation cost. To consider the impact of point defects including $\mathrm{V_O}$, $\mathrm{V_{Th}}$, $\mathrm{I_O}$, and $\mathrm{I_{Th}}$ on phonon transport, defect-bearing supercells are constructed by removing or adding Th or O atoms in the atomic configurations. We create vacancies by removing O or Th atoms and create interstitials by inserting O or Th atoms at the octahedral site. Three concentrations of point defects, 0.067\%, 0.133\%, and 0.267\%, are achieved by removing or adding 1, 2, and 4 Th or O atoms, respectively. Defects are even distributed in the simulation cells to avoid strong interactions.

\begin{figure}[htbp]
    \centering
    \includegraphics[width=0.8\textwidth]{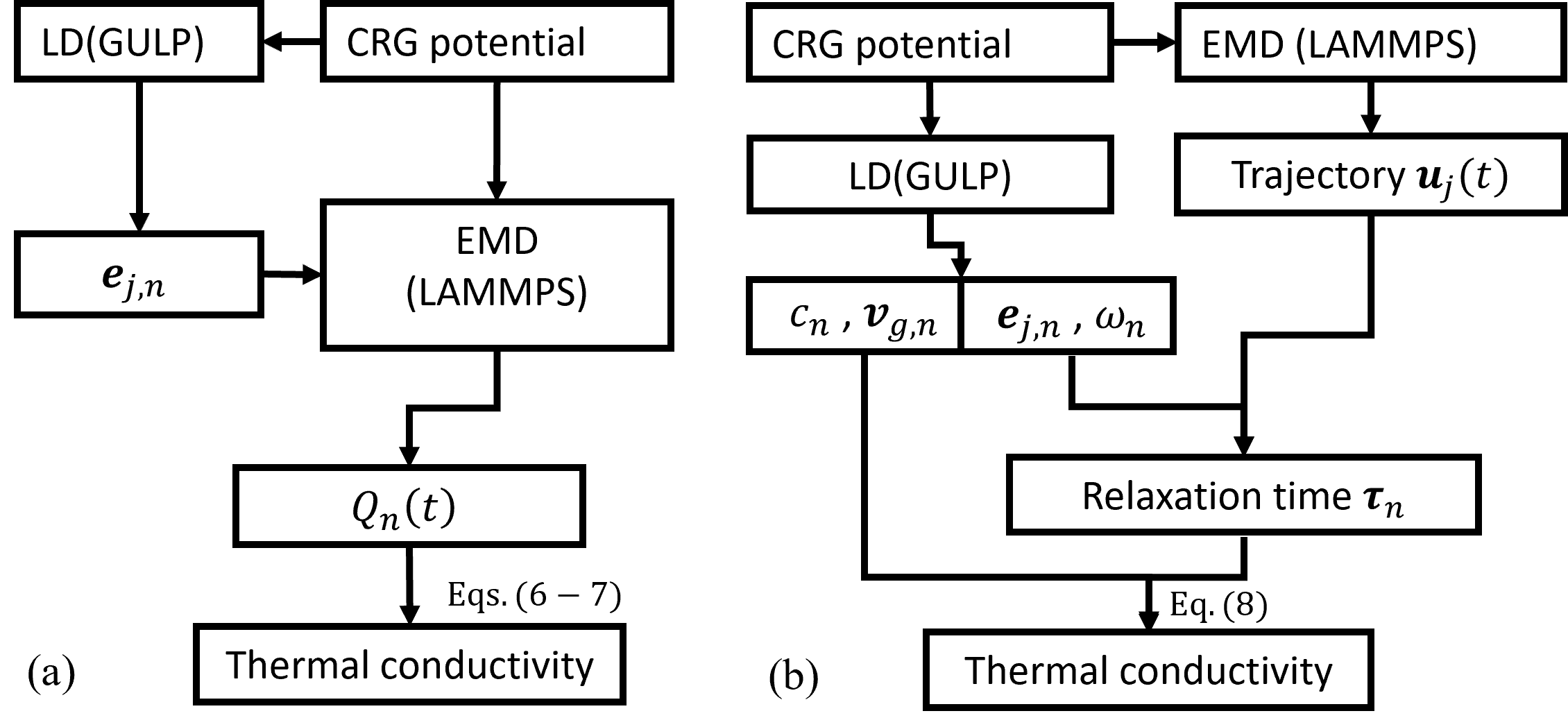}
    \caption{Flow chart of (a) GKMA method and (b) RTA-NMA method.}
    \label{fig:1}
\end{figure}

The procedure is shown in FIG. \ref{fig:1}.  In GKMA (FIG. \ref{fig:1}a), after the same relaxation procedure for the supercell, the eigenvectors are incorporated in the equilibrium process and the phonon modal flux (Eq. (\ref{eq_gkma_heatflux})) is calculated on the fly to reduce computation cost \cite{Seyf2019}. Then the phonon modal thermal conductivity can be calculated using the GK formula. Periodic boundary conditions are set up in all MD simulations. To ensure statistical quality, 100 independent simulations from different random seeds are performed for each temperature and point defect combination. In RTA-NMA (FIG. \ref{fig:1}b), $\mathbf{v}_{g,n}$ and $\mathbf{e}_{j,n}$ are derived from LD of the defect-free supercell, and the atomic trajectory $u_{j}(t)$ is obtained from EMD simulation. For EMD, we first thermalize the supercell structure for 200 ps at a specific temperature under the canonical ensemble (NVT). Here, thermal expansion is ignored due to the low-temperature range ($<0.2 ~T_m$ where $T_m$ is the melting point) considered in this study. Then, the atomic trajectory is recorded for 400 ns under the microcanonical ensemble (NVE) using the functionality available in the LAMMPS package \cite{LAMMPS}. Such trajectory information is post-analyzed, considering phonon eigenvectors and group velocities.

As a first approximation, we will quantify the change in thermal conductivity due to temperature and defects via Eq. (\ref{eq_abc}) \cite{Bakker1997,Liu2016,Hurley2022}, where $T$ is temperature, $x$ is the fraction of point defects, $b$ (in unit m/W) describes the intrinsic phonon-phonon scattering and theoretically remains constant, $c$ (in unit mK/W) describes phonon-defect scattering, and $a$ (in unit mK/W) describes all other effects that affect thermal transport. The temperature range is set at 250-700 K to avoid defect migration during the simulation process, which can complicate the phonon analysis.

\begin{equation}\label{eq_abc}
 \kappa = \frac{1}{a+bT+cx}
\end{equation}

\section{Results}

\subsection{GKMA}

\begin{figure}[htbp]
    \centering
    \includegraphics[width=0.55\textwidth]{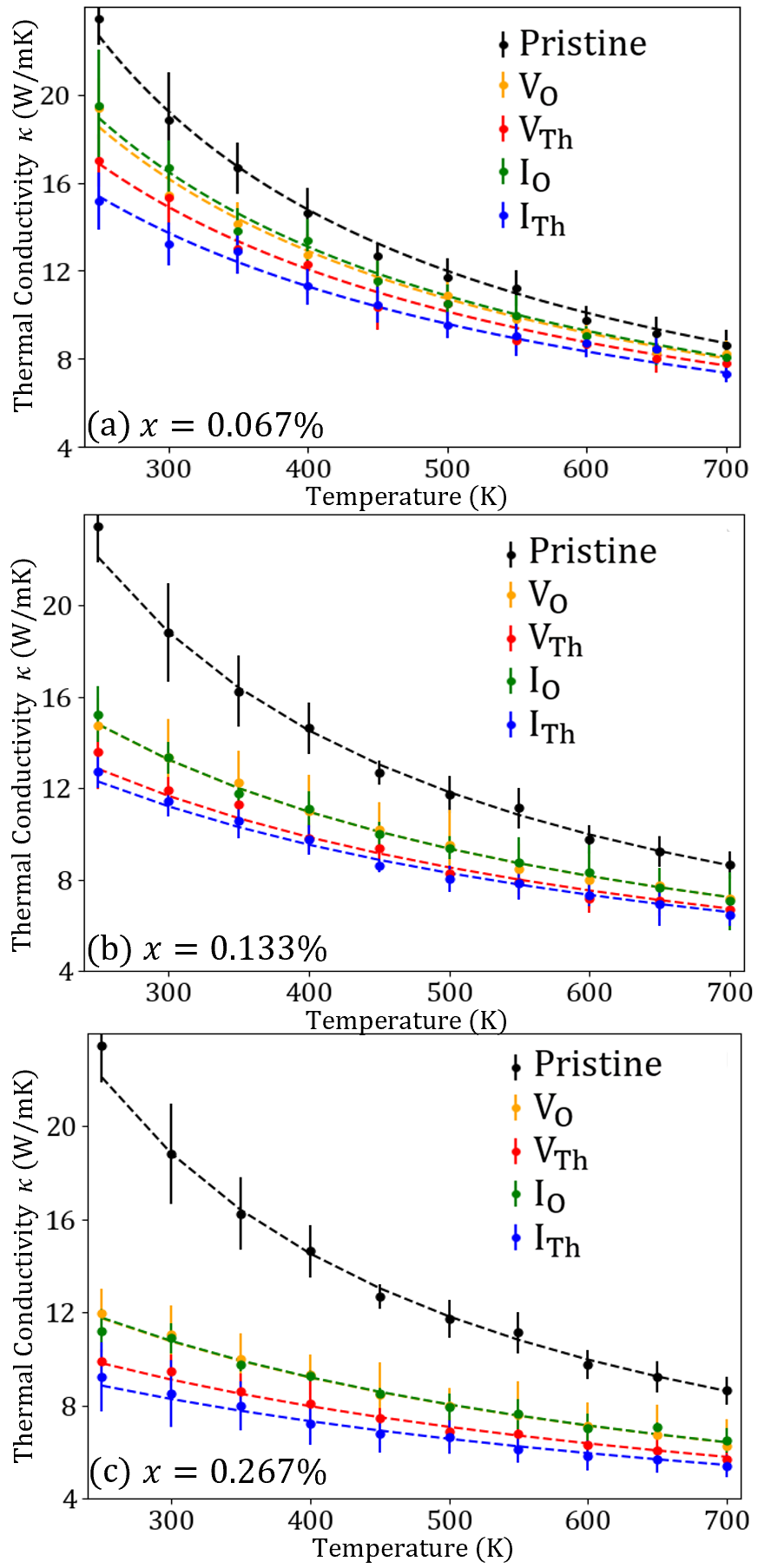}
    \caption{Results of thermal conductivity from GKMA method as a function of temperature (ranging from 250 K to 700 K with 50 K increment) for pristine ThO$_2$ and ThO$_2$ with different concentrations of $\mathrm{V_O}$, $\mathrm{V_{Th}}$, $\mathrm{I_O}$ and $\mathrm{I_{Th}}$: (a) 0.067\%, (b) 0.133\%, and (c) 0.267\%. The dashed lines are fitted curves according to Eq. (\ref{eq_abc}). }
    \label{fig2gktc}
\end{figure}

Based on the GKMA method, FIGs. \ref{fig2gktc}a-c summarize the thermal conductivity as a function of temperature for pristine ThO$_2$ and ThO$_2$ with point defects at concentrations of 0.067\%, 0.133\%, and 0.267\%, respectively. The results indicate that the thermal conductivity degradation due to point defects depends on the type and concentration of the defects. Among the considered defects, $\mathrm{I_{Th}}$ leads to the largest degradation, followed by $\mathrm{V_{Th}}$, and then $\mathrm{I_O}$ and $\mathrm{V_O}$ show the similar impact. As expected, a larger concentration of the defects led to lower thermal conductivity.

\begin{figure}[htbp]
    \centering
    \includegraphics[width=0.55\textwidth]{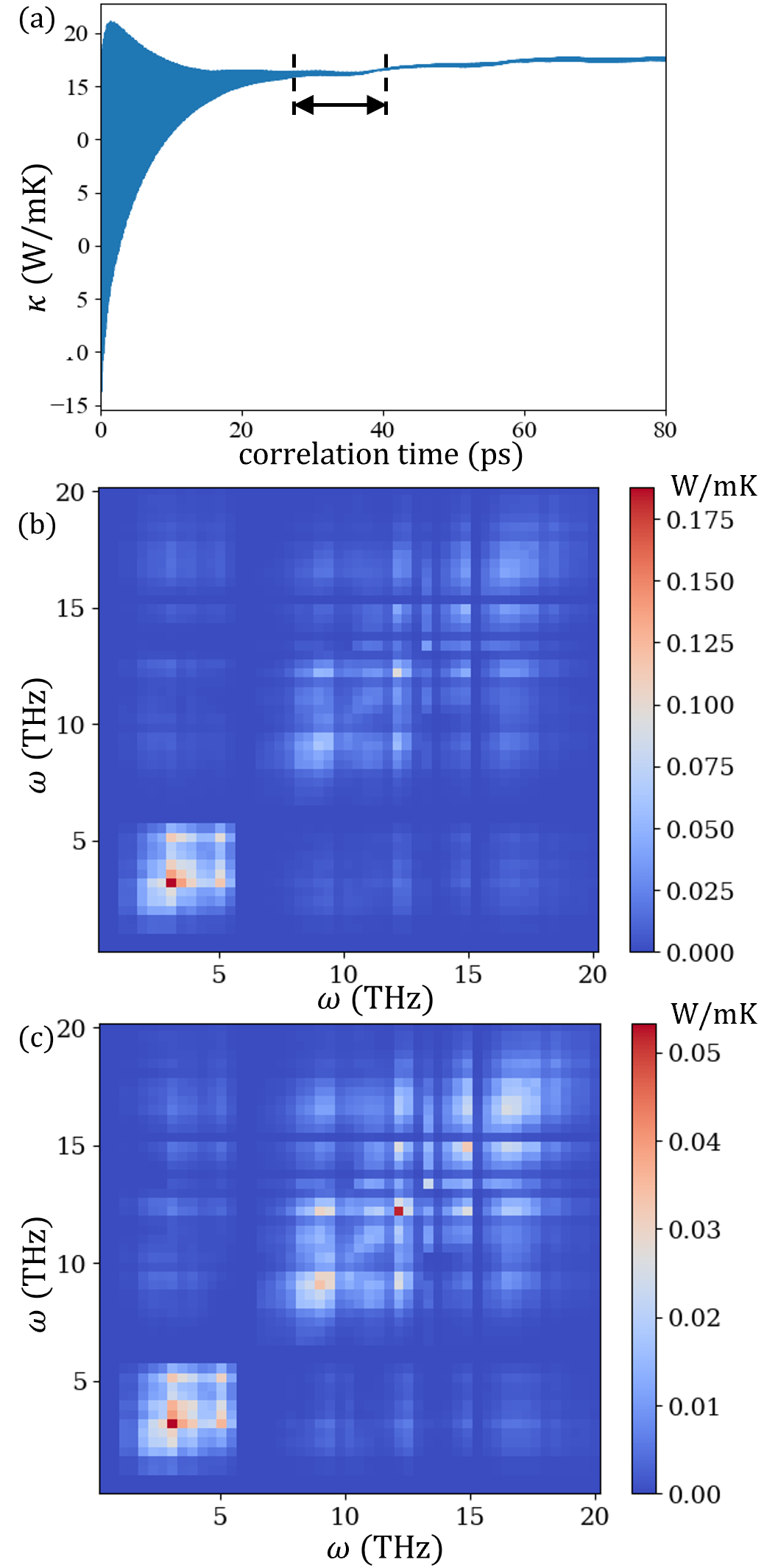}
    \caption{(a) Integration of heat flux correlation, $\kappa$ is calculated using the period marked by double arrows. (b-c) 2D phonon spectral thermal conductivity map (b) for pristine ThO$_2$ and (c) for ThO$_2$ with $\mathrm{I_{Th}}$ at a concentration of 0.267\% at 300 K, respectively.} 
    \label{fig3gkmap}
\end{figure}

FIG. \ref{fig3gkmap}a provides one typical example of integrated thermal conductivity versus auto-correlation time based on the GK formula. The temporal evolution of thermal conductivity features initial strong fluctuations and gradually stabilizes over time. Hence, we use the average values as marked between the double arrows in FIG. \ref{fig3gkmap}a to reduce statistical errors. FIGs. \ref{fig3gkmap}b-c show the cross-modal conductivity results calculated from  Eq. (\ref{eq_gkma_2d})  for pristine ThO$_2$ and ThO$_2$ with $\mathrm{I_{Th}}$. A significant difference in the thermal conductivity distribution across phonon frequencies is observed between Fig. 3b and 3c. Notably, with the presence of $\mathrm{I_{Th}}$, the relative contribution from the optical phonon modes ($\omega_n >$ 6 THz based on \cite{Jin2021}) increases, which indicates that the relative acoustic phonon mode contribution to thermal conductivity is markedly reduced. Additional 2D thermal conductivity maps for $\mathrm{V_O}$, $\mathrm{V_{Th}}$, $\mathrm{O_i}$ are provided in SM Figure S2-S4.

\subsection{RTA-NMA}

\begin{figure}[htbp]
    \includegraphics[width=\textwidth]{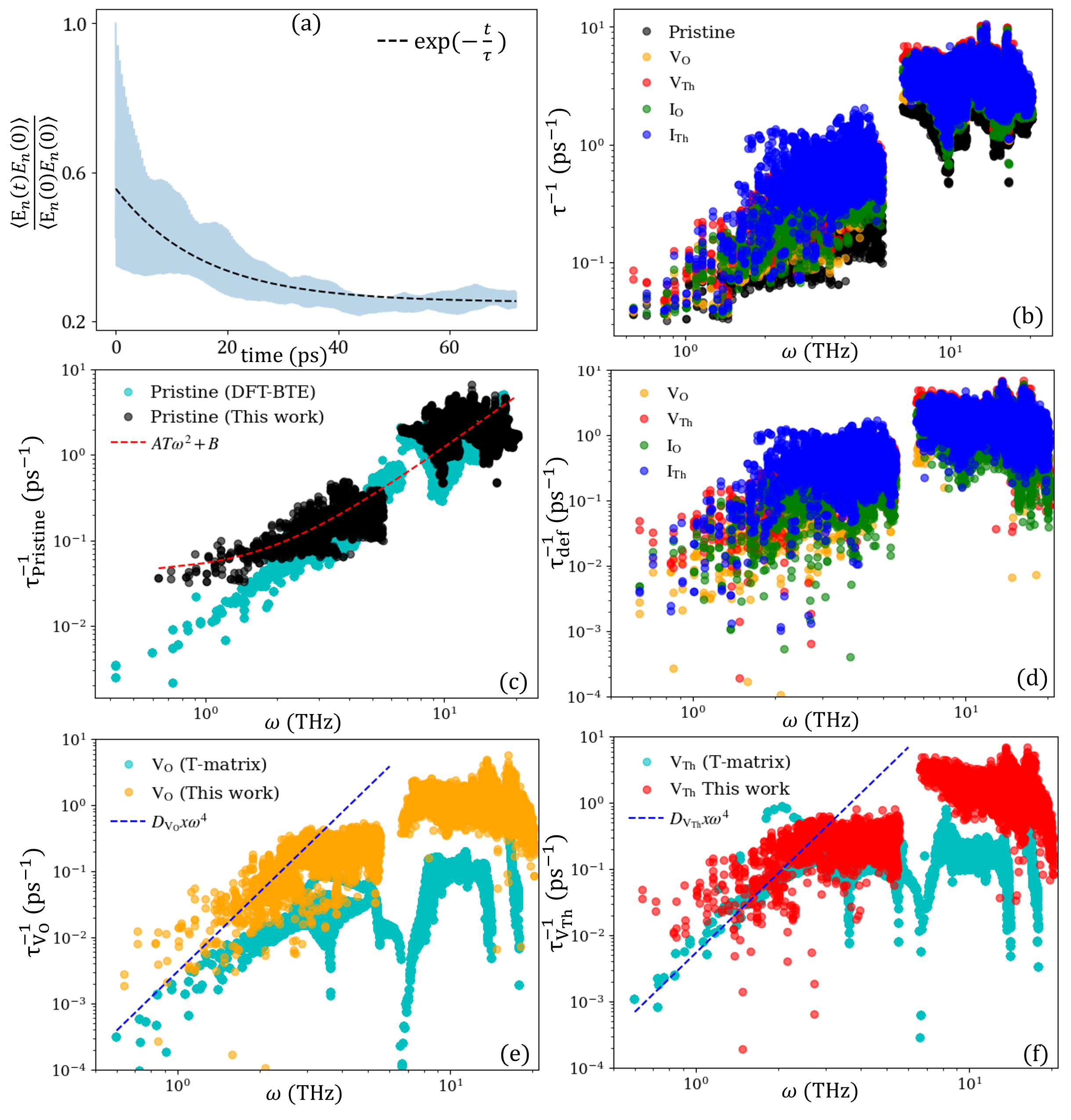}
    \centering
	\caption{(a) An example of using Eq. (\ref{eq_nma_ac}) to obtain $\tau$ for a phonon mode with $\omega_n=1.45$ THz. (b) $\tau^{-1}$  of pristine ThO$_2$ and ThO$_2$ with point defects ($\mathrm{V_O}$, $\mathrm{V_{Th}}$, $\mathrm{I_O}$ and $\mathrm{I_{Th}}$) at a concentration of 0.267\%  at 300 K (c) $\tau^{-1}_{\mathrm{Pristine}}$ comparison at 300 K between this work and Malakkal et al.'s data \cite{Malakkal2024} (d) $\tau^{-1}_\mathrm{defect}$ for $\mathrm{V_O}$, $\mathrm{V_{Th}}$, $\mathrm{I_O}$ and $\mathrm{I_{Th}}$ at a concentration of 0.267\% at 300 K. (e-f)$\tau^{-1}_\mathrm{defect}$ comparison at 300 K between this work and Malakkal et al.'s data for (e) $\mathrm{V_O}$ and (f) $\mathrm{V_{Th}}$ at a concentration of 0.267\%. Dash lines fitted from the Klemens-Callaway model are shown in (c), (e), and (f).}
    \label{fig4tau} 
\end{figure}

In the RTA-NMA method, we first obtain the phonon relaxation time ($\tau$). FIG. \ref{fig4tau}a illustrates an example where $\tau_n$ of an acoustic phonon mode ($\omega_n$=1.45 THz) is obtained by exponential fitting of the normalized auto-correlation function of phonon mode total energy using Eq. (\ref{eq_nma_ac}). To ensure the accuracy of relaxation time values, the fitting results are averaged over 100 independent simulations for each phonon mode (see details in SM, part 1). The overall phonon scattering rates ($\tau^{-1}$, the inverse of the phonon relaxation time) at 300 K for pristine ThO$_2$ and ThO$_2$ with point defects ($\mathrm{V_O}$, $\mathrm{V_{Th}}$, $\mathrm{I_O}$ and $\mathrm{I_{Th}}$ at a concentration of 0.267\%) are calculated and plotted in FIG. \ref{fig4tau}b. 


According to the Klemens-Callaway model \cite{klemens1958thermal, Callaway1959, Liu2016}, in this work, the overall phonon scattering rate can be written as

\begin{equation}
\begin{aligned}
    \tau^{-1}& = \tau^{-1}_{\mathrm{ph}} + \tau^{-1}_{\mathrm{B}} + \tau^{-1}_{\mathrm{def}}=AT\omega^2 + B + D_\mathrm{def}x\omega^4, \\
    \tau_{\mathrm{pristine}} &=  \tau^{-1}_{\mathrm{ph}} + \tau^{-1}_{\mathrm{B}} = AT\omega^2 + B, \\ \tau^{-1}_{\mathrm{def}}&=D_\mathrm{def}x\omega^4,    \\
\end{aligned}
\end{equation}

where  $\tau^{-1}_{\mathrm{ph}}=AT\omega^2$ is intrinsic phonon-phonon scattering rate (T is temperature), $\tau^{-1}_{\mathrm{B}}=B$ is phonon-boundary scattering rate, and $\tau^{-1}_{\mathrm{def}}$ is the phonon-defect scattering rate and the defect can be the vacancy ($\mathrm{V_O}$ or $\mathrm{V_{Th}}$) or interstitial ($\mathrm{I_O}$ or $\mathrm{I_{Th}}$). $A$, $B$ are fitted by $\tau^{-1}_{\mathrm{pristine}}$ phonon scattering rate of pristine ThO$_2$, and $D_\mathrm{def}$ can be fitted by $\tau^{-1}_{\mathrm{def}}=\tau^{-1} -\tau^{-1}_{\mathrm{pristine}} $, and their values are shown in Table \ref{tab:callaway}.
\begin{table}[htbp]
\small
  \caption{Fitted values of $A$, $B$, $D_{\mathrm{V_O}}$, $D_{\mathrm{V_{Th}}}$, $D_{\mathrm{I_O}}$ and $D_{\mathrm{I_{Th}}}$.}
  \begin{tabular*}{\textwidth}{@{\extracolsep{\fill}}llllll}
    \hline
    $A(\mathrm{ps\cdot K^{-1}})$ & $B(\mathrm{ps}^{-1})$ & $D_{\mathrm{V_O}}(\mathrm{ps}^{-3})$ & $D_{\mathrm{V_{Th}}}(\mathrm{ps}^{-3})$ & $D_{\mathrm{I_O}}(\mathrm{ps}^{-3})$ &$D_{\mathrm{I_{Th}}}(\mathrm{ps}^{-3})$\\
    \hline 
    $4.09\times10^{-5}$ & $4.96\times10^{-2}$ & $1.16$ & $2.11$ & $1.46$ & $2.84$ \\
    \hline
  \end{tabular*}
  \label{tab:callaway}
\end{table}

In FIG. \ref{fig4tau}c, the phonon scattering rate of pristine ThO$_2$ at 300 K, are compared with data calculated from Malakkal et al. \cite{Malakkal2024} using perturbative approaches based on density function theory (DFT) with BTE (the DFT-BTE \cite{Li2014} method). The data from the two methods overlaps at most phonon frequencies except in the low-frequency range where DFT-BTE predicts lower scattering rates. This discrepancy is partly attributed to higher-order anharmonicity effects that can not be captured by the DFT-BTE method, which uses only third-order IFCs. Another factor to consider is that the computation by Malakkal et al. is at the ground state without considering finite temperature dynamics. In addition, we note there is a distinct gap at around 6 THz between acoustic ($\omega<6$ THz) and optical ($\omega>6$ THz) phonons, which is not apparent from DFT data; this is due to the inconsistency in band gap predictions \cite{Jin2021}.

In FIG. \ref{fig4tau}d, the phonon-defect scattering rates $\tau^{-1}{\mathrm{def}}$ at 300 K for $\mathrm{V_O}$, $\mathrm{V_{Th}}$, $\mathrm{I_O}$, and $\mathrm{I_{Th}}$ at a concentration of 0.267\% are shown (results for other concentrations indicate a consistent trend). The results demonstrate that Th defects exhibit higher phonon-defect scattering rates than O defects. In FIG. \ref{fig4tau}e-f,  $\tau^{-1}_{\mathrm{V_O}}$ and $\tau^{-1}_{\mathrm{V_{Th}}}$ at a concentration of 0.267\% are compared with data calculated by the T-matrix method \cite{Kundu2011} from Malakkal et al. \cite{Malakkal2024} at 300 K. For $\mathrm{V_O}$,  $\tau^{-1}_{\mathrm{V_O}}=D_\mathrm{V_O}x\omega^4$ captures the acoustic phonon scattering rate well at $\omega <$ 3 THz and the T-matrix method predicts lower phonon-defect scattering rate at almost the whole frequency range. For $\mathrm{V_{Th}}$, the deviation of $\tau^{-1}_{\mathrm{V_{Th}}}$ is larger but fitted $D_\mathrm{V_{Th}}x\omega^4$ curve shows good agreement with T-matrix data.

\begin{figure}[htbp]
    \centering
    \includegraphics[width=0.55\textwidth]{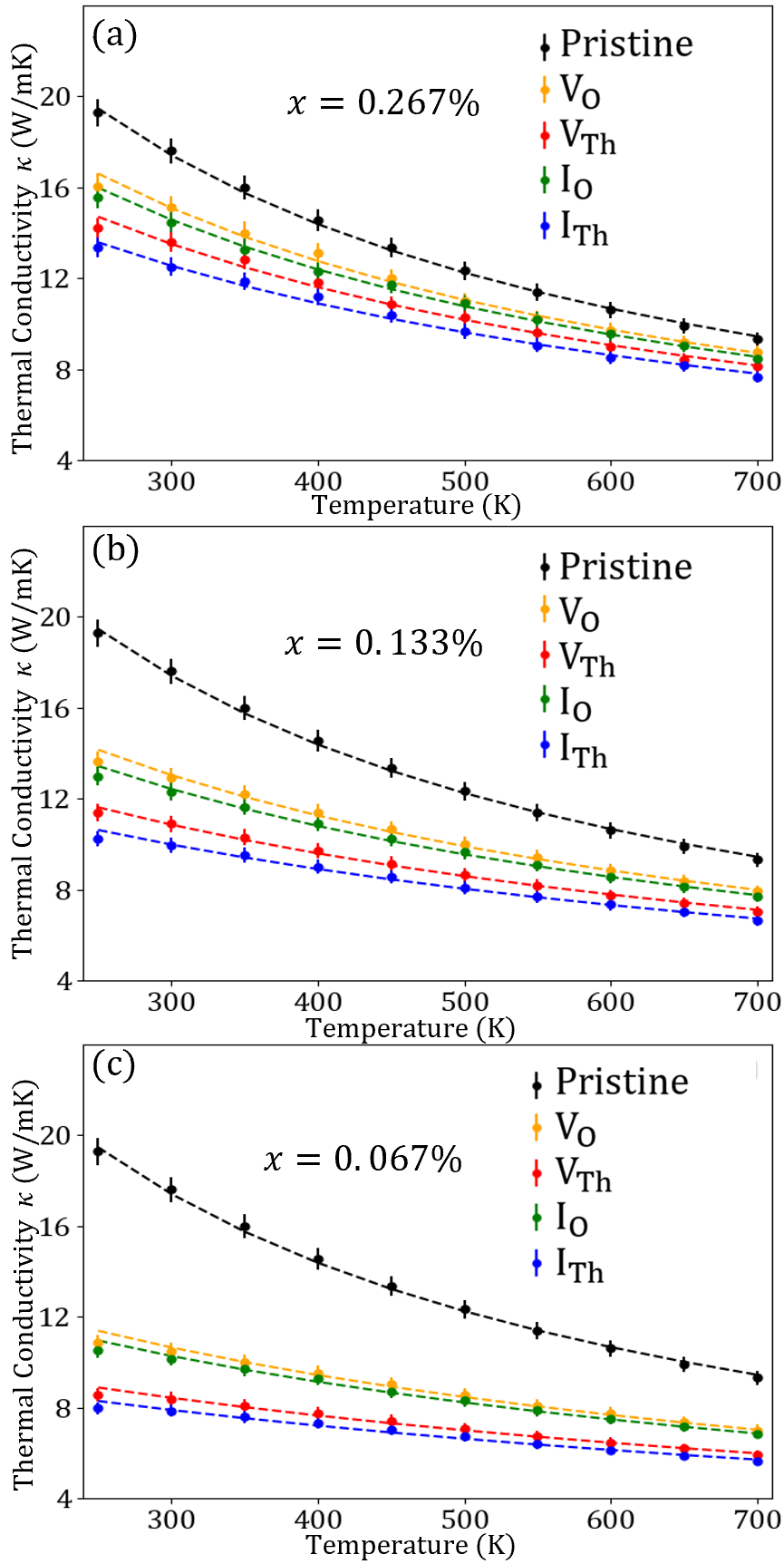}
    \caption{Results of thermal conductivity from RTA-NMA method as a function of temperature (ranging from 250 K to 700 K with 50 K increment) for pristine ThO$_2$ and ThO$_2$ with different concentrations of $\mathrm{V_O}$, $\mathrm{V_{Th}}$, $\mathrm{I_O}$ and $\mathrm{I_{Th}}$: (a) 0.067\%, (b) 0.133\%  and (c) 0.267\%. The dashed lines are fitted curves according to Eq. (\ref{eq_abc}).} 
 \label{fig5rtatc}
\end{figure}

FIG. \ref{fig5rtatc} summarizes the thermal conductivity calculated using the Eq. (\ref{eq_nma_kappa}) for pristine ThO$_2$ and ThO$_2$ with different point defects at varying concentrations. These values are in reasonable agreement with the results obtained from the GKMA method. The results reveal that thermal conductivity degradation by the considered intrinsic defect followed the same pattern as predicted by the GKMA method, i.e, $\mathrm{I_{Th}}>\mathrm{V_{Th}}>\mathrm{I_O}\approx\mathrm{V_O}$. However, a slower decline of thermal conductivity with increasing temperature is identified.

\begin{figure}[htbp]
    \centering
    \includegraphics[width=0.55\textwidth]{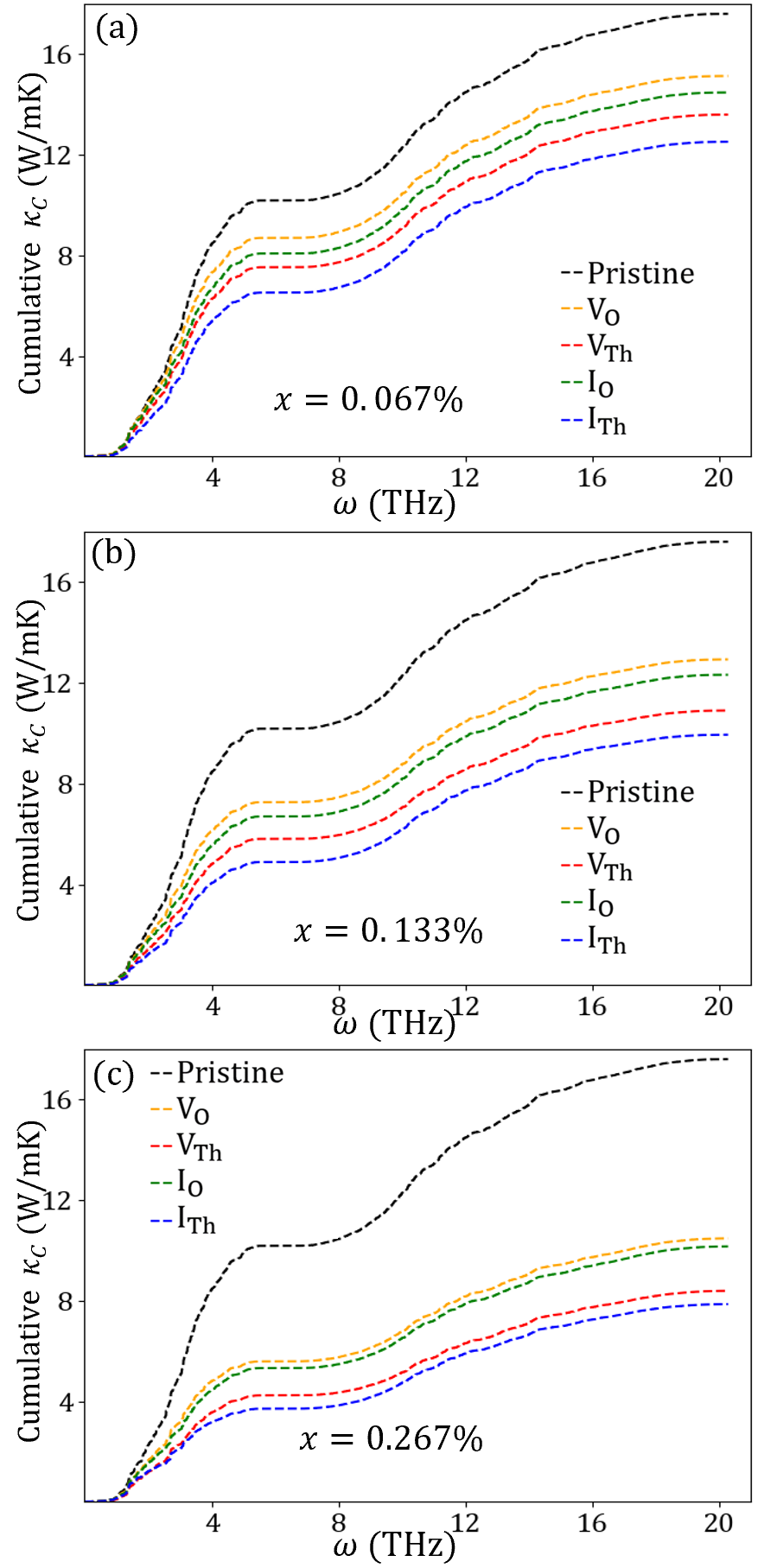}
    \caption{Cumulative thermal conductivity of pristine ThO$_2$ and ThO$_2$ with different point defects ($\mathrm{V_O}$, $\mathrm{V_{Th}}$, $\mathrm{I_O}$ and $\mathrm{I_{Th}}$) at different concentrations of (a) 0.067\% , (b) 0.133\% and (c) 0.267\%  at 300 K by RTA-NMA method.} 
    \label{fig6rtacu}
\end{figure}

The cumulative thermal conductivity as a function of phonon frequency at 300 K is also determined using Eq. (\ref{eq_nma_cumulative}), shown in FIG. \ref{fig6rtacu}, which indicates that the thermal conductivity of ThO$_2$ is mainly contributed by acoustic phonon modes, consistent with previous studies \cite{Jin2021,Malakkal2019}; the presence of point defects leads to a decrease in the acoustic phonon contribution to thermal conductivity (see SM Table S1 for both 300 K and 700 K results, where the latter leads to a slight reduction in acoustic phonon contribution). It can be seen that $\mathrm{I_{Th}}$ carries the most reduction in the acoustic contribution, which is consistent with the largest acoustic phonon-defect scattering rates observed for $\mathrm{I_{Th}}$, as illustrated in FIG. \ref{fig4tau}d.

\subsection{Comparison between GKMA and RTA-NMA methods}

From the absolute values of thermal conductivity in FIG. \ref{fig2gktc} and  FIG. \ref{fig5rtatc}, it can be generally concluded that the RTA-NMA method leads to larger values at higher temperatures than that from the GKMA method. Both methods incorporate phonon eigenvectors to calculate certain quantities. Particularly, the GKMA method does not need phonon-specific properties such as group velocity and specific heat; instead, the thermal conductivity is directly calculated from the auto-correlation of heat flux, which does not rely on the relaxation time approximation as in the RTA-NMA method \cite{Henry2009}. Hence, although the GKMA method may encounter statistical variations in its implementation, it can, in principle, provide more realistic estimations for thermal conductivity at elevated temperatures.

To evaluate the trend in the thermal conductivity as a function of temperature, results from both methods are fitted using Eq. (\ref{eq_abc}). For GKMA method, $a=4.8\times10^{-3}$ mK/W and $b=1.57\times10^{-4}$ m/W; for RTA-NMA method, $a=1.88\times10^{-2}$ mK/W and $b=1.25\times10^{-4}$ m/W. The $b$ values for the pristine ThO$_2$ reported by the previous studies are scattered. For example, simulation work yielded 1.70$\mathrm{\times10^{-4}}$ m/W by Jin et al. \cite{Jin2022}, and 2.19$\mathrm{\times10^{-4}}$ m/W by Cooper et al. \cite{COOPER201529}, while experiment led to 2.25$\mathrm{\times10^{-4}}$ m/W by Bakker et al.  \cite{Bakker1997} and 1.98$\mathrm{\times10^{-4}}$ m/W by Malakkal et al. \cite{Malakkal2019}. In the current work, GKMA exhibits a better agreement with previous work, which suggests that, compared to GKMA, RTA-NMA tends to overestimate thermal conductivity at high temperatures (See SM Table S2 for a numerical comparison of thermal conductivity values).

\begin{table}[htbp]
\small
  \caption{Fitted values of $c$ for $\mathrm{V_O}$, $\mathrm{V_{Th}}$, $\mathrm{I_O}$ and $\mathrm{I_{Th}}$ by RTA-NMA and GKMA methods, compared to results from Jin et al. \cite{Jin2022}.}
  \begin{tabular*}{\textwidth}{p{1.8cm} @{\extracolsep{\fill}} *{3}{p{4cm}}}
    \hline
    $c$ (mK/W)  & RTA-NMA & GKMA&  Jin et al. \cite{Jin2022}\\
    \hline 
     & \multicolumn{2}{p{8cm}}{($x$ range from $0.067\%$ to $0.267\%$ in this work)}  & ($x$ $0.2\%$ to $1\%$)\\
    \hline 
    $c_\mathrm{V_O}$   & 14.96 & 15.92  & 7.32\\
    $c_\mathrm{V_{Th}}$ & 25.36 &  23.26  & 13.41\\
    $c_\mathrm{I_O}$  & 17.19 & 15.34 & 8.38\\
    $c_\mathrm{I_{Th}}$  & 31.23 & 28.38 & 16.60\\
    \hline
  \end{tabular*}
  \label{tab:4}
\end{table}

With the consideration of defects, the $c$ values from the two methods are highly consistent (Table \ref{tab:4}), which indicates the strongest impact of $\mathrm{I_{Th}}$, followed by $\mathrm{V_{Th}}$, and then oxygen defects. Note that these values are much larger than the values reported by Jin et al. \cite{Jin2022}, i.e., that the degradation of thermal conductivity with increasing defect concentration is larger in this work. This difference is likely due to the different defect concentration ranges chosen for fitting the curve, where in Jin et al.'s work \cite{Jin2022}, the defect concentration is much higher (0.2\% to 1\%). As reported by Jin et al. \cite{Jin2022} and Dennett et al. \cite{Dennett2021}, the rate of conductivity degradation due to point defect is more significant at low defect concentrations and slows down with higher defect concentrations. 
From the difference, it can be inferred that the linear relationship between $1/\kappa$ versus $c$ as described by Eq. (\ref{eq_abc}) might be too simplified, particularly in the situations of high defect concentrations where the reduced distance between defects could lead to enhanced defect interactions via elastic and Coulombic interactions.

\section{Discussion}

\begin{figure}
    \centering
    \includegraphics[width=0.6\linewidth]{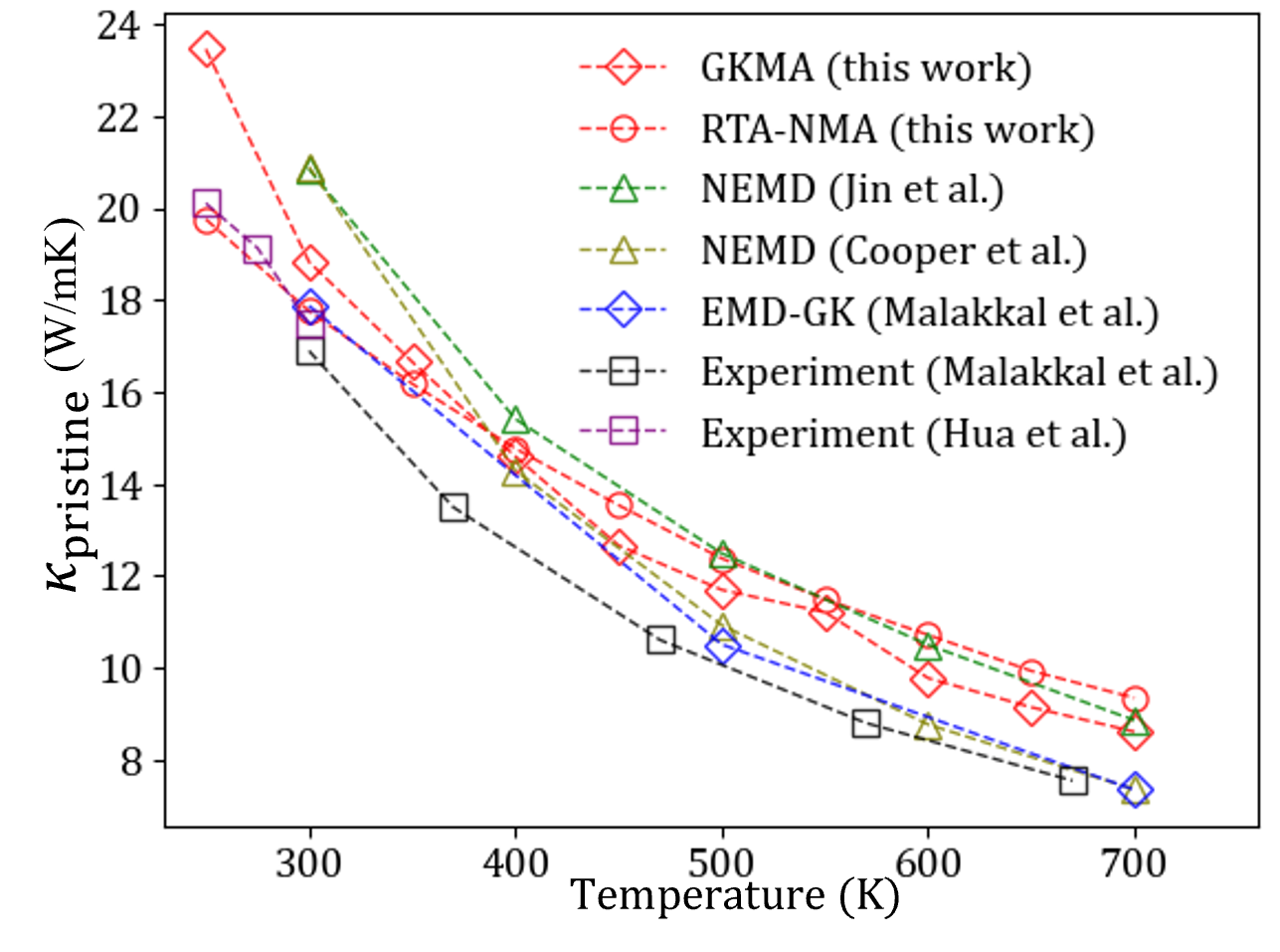}
    \caption{Thermal conductivity versus temperature for pristine ThO$_2$ from this work are compared with Jin et al. \cite{Jin2022}, Cooper et al. \cite{COOPER201529}, Malakkal et al. \cite{Malakkal2019} and Hua et al. \cite{HUA2023}.}
    \label{fig_kappas}
\end{figure}

As a validation, the thermal conductivity results for pristine ThO$_2$ are compared with the literature. FIG. \ref{fig_kappas} compiles thermal conductivity from two methods in this work, EMD simulations by Malakkal et al. \cite{Malakkal2019}, NEMD simulations by Jin et al. \cite{Jin2022} and Cooper et al. \cite{COOPER201529}, and experiments by Hua et al. \cite{HUA2023} and Malakkal et al. \cite{Malakkal2019}. These data exhibit consistent trends at the temperature range from 250 K to 700 K. From experiments, Malakkal et al. \cite{Malakkal2019} measured conductivity over a broad range of temperatures. At 300 K,  slightly lower values were found than Hua et al. \cite{HUA2023}. This is likely due to the sample quality as Hua et al. used high-quality single crystal ThO$_2$ while Malakkal et al. used spark plasma sintered ceramic ThO$_2$ and the measured values were scaled to 100\% theoretical density. In the current work, the thermal conductivity of pristine ThO$_2$ at 250 K and 300 K from RTA-NMA is 19.75 W/mK and 17.76 W/mK, respectively, which are in strong agreement with Hua et al.'s measurements at 20.10 W/mK and 17.48 W/mK for the two temperatures. 

Based on NEMD simulations, Jin et al. (simulation box length 70-170 nm) \cite{Jin2022} and Cooper et al. (simulation box length 10-30 nm) \cite{COOPER201529} report thermal conductivity of 20.84 W/mK and 20.89 W/mK at 300 K, respectively, by extrapolating their data based on multiple simulation cell lengths to estimate thermal conductivity values at infinite cell length, effectively eliminating size dependencies. These values are higher than EMD simulation results from this work (simulation box size 2.8 nm), Malakkal et al. (simulation box size 4.5-11 nm) \cite{Malakkal2019} with 17.87 W/mK, and Ma et al. (simulation box size 3.4 nm) \cite{MA2015476} with 16.67 W/mK at 300 K. With increasing temperature, the difference becomes less consistent. This discrepancy between NEMD and EMD results is partly attributed to the treatment of the NEMD finite-size effect, where linear extrapolation is often performed to obtain conductivity at infinite supercell length due to the numerical setup of energy reservoirs that scatter phonons \cite{COOPER201529,Jin2022}.

The substantial contribution of acoustic phonons to the thermal conductivity of pristine ThO$_2$ at 300 K has been discussed in previous studies \cite{Jin2021,Malakkal2024,LIU201811}. The proportion of acoustic contribution is identified to be around 73\% for pristine ThO$_2$ by Liu et al. \cite{LIU201811} and  Malakkal et al. \cite{Malakkal2024} based on the second and third-order IFCs derived from DFT calculations. Further, Jin et al. \cite{Jin2021} reported an acoustic contribution of 67.5\% using IFCs based on the CRG interatomic potential as used in this work. However, here, we note that the acoustic contribution is 60\% from the RTA-NMA method. The distinction of this work is the incorporation of dynamics in computing phonon relaxation time, which takes full consideration of phonon scattering at all orders from the dynamics at finite temperatures. This comparison suggests that the computation of relaxation time that relies only on three-phonon processes accounts for around 7.5\% overestimation of the acoustic contribution. 

\begin{table}[htbp]
\caption{ $\mathrm{ \kappa_{defect}/\kappa_{pristine}}$ for point defects of a concentration of 0.133\% at 300 K. }
\label{tbl:5}
\begin{tabular*}{\textwidth}{@{\extracolsep{\fill}}llllll}
\hline
      & RTA-NMA& GKMA & Deskins et al. \cite{Deskins2021} &Park et al. \cite{Park2018defect}  & Malakkal et al. \cite{Malakkal2024}\\ \hline
$\mathrm{V_O}$  &0.72 &0.71 &0.67 &0.73 & 0.87        \\ 
$\mathrm{V_{Th}}$ &0.64 &0.63 &0.59 &0.60  & 0.65    \\ 
$\mathrm{I_O}$  &0.70 & 0.71 & 0.73               \\ 
$\mathrm{I_{Th}}$ &0.57 & 0.61 & 0.59             \\ 
\hline
\end{tabular*}%

\end{table}

Finally, we discuss the prediction of degradation in thermal conductivity based on different theoretical methods regarding the individual point defect at the same concentration at 300 K. Table \ref{tbl:5} summarizes the current results and the work of Park et al. \cite{Park2018defect} using NEMD, Deskins et al. \cite{Deskins2021} using Klemens model, and Malakkal et al. \cite{Malakkal2024} using the T-matrix method. Note that in both the Klemens model and T-matrix calculations, there is no consideration of finite temperature dynamics. The current work indicates the impact as $\mathrm{I_{Th}}>\mathrm{V_{Th}}>\mathrm{I_O}\approx\mathrm{V_O}$. Particularly, the results of $\mathrm{V_{Th}}$ show a good consistency across all methods. For $\mathrm{V_O}$, by contrast, Malakkal et al.'s work \cite{Malakkal2024} indicates a much weaker degradation than others. This is consistent with Fig.\ref{fig4tau}e-f where lower phonon-defect scattering rates lead to longer phonon relaxation time and thus a higher estimation of the thermal conductivity with $\mathrm{V_O}$. As for interstitials, $\mathrm{I_{Th}}$ is obviously a stronger phonon scattering center than $\mathrm{I_O}$, based on the current work and Deskins et al. \cite{Deskins2021}. 

\section{Conclusion}

In this work, thermal transport in pristine ThO$_2$ and ThO$_2$ with point defects ($\mathrm{V_O}$, $\mathrm{V_{Th}}$, $\mathrm{I_O}$ and $\mathrm{I_{Th}}$) at concentrations of 0.067\%, 0.133\%, and 0.267\% is evaluated from 250 K to 700 K. Both RTA-NMA and GKMA methods based on LD and EMD are employed to assess the phonon modal contribution to the thermal conductivity, which enable different perspectives of how point defects affect phonon transport, such as phonon relaxation times under finite temperature dynamics and cross-modal contributions to conductivity. Extensive comparisons are made with other theoretical methods. Among the four types of point defects, the $\mathrm{I_{Th}}$  exerts the most influence, followed by the $\mathrm{V_{Th}}$, and then the $\mathrm{V_O}$ and $\mathrm{I_O}$ which exhibit a comparable level of impact. These analyses also highlight the substantial contribution of acoustic phonons to thermal conductivity. The degradation in thermal conductivity caused by point defects is primarily attributed to the enhanced scattering of acoustic phonons interacting with these defects.

\section*{Acknowledgments}
This work is supported by the Center for Thermal Energy Transport under Irradiation, an Energy Frontier Research Center funded by the U.S. Department of Energy, Office of Science, United States, Office of Basic Energy Sciences. 

\bibliographystyle{elsarticle-num-names} 

\bibliography{manuscript.bib}

\end{document}


\maketitle

\section{Average value and error bar}

Molecular Dynamics has intrinsic statistical error, so to get reliable data, for every system at every temperature, 10 simulations are used to calculate the average to get a value, and 10 values are used to calculate an averaged value and standard error. This implies that for each thermal conductivity data point, we conducted 100 simulations using LAMMPS, all with identical setups but different random seeds.

In NMA, for every 10 runs, 10 sets of $\tau$ are obtained from exponential fitting, and then by 

\begin{equation}
    1/\tau_{average} = (1/\tau_{1} +1/\tau_{2} +......+1/\tau_{10})/10
\end{equation}

phonon lifetime is calculated from these 10 outputs, and then the thermal conductivity value is calculated from phonon lifetimes.
After 10 values of thermal conductivity are calculated, the mean value and standard error are conducted.

In GKMA, for every 10 runs, 10 sets of $\langle Q_{m}(t)Q_{m^{'}}(0)\rangle$ was obtained, and then by

\begin{equation}
    \langle Q_{m}(t)Q_{m^{'}}(0)\rangle_{average}=(\sum^{10}_{i=1}  \langle Q_{m}(t)Q_{m^{'}}(0)\rangle_{i}) / 10
\end{equation}

Then $\langle Q_{m}(t)Q_{m^{'}}(0)\rangle_{average}$ was used to calculate the thermal conductivity result of these 10 runs.

\section{Tables and figures}

\begin{table}[htpb]
\centering
\caption{The acoustic phonon mode contribution at 300 K and 700 K in (\%). }
\label{tab:acoustic_contribution}

\begin{tabular*}{\textwidth}{@{\extracolsep{\fill}}llllllll}
\hline
$x$(\%)    & \multicolumn{2}{c}{0.067} & \multicolumn{2}{c}{0.133} & \multicolumn{2}{c}{0.267} \\
\hline
$T$(K)     & 300         & 700         & 300         & 700         & 300         & 700         \\
\hline
Pristine & 59          & 57          & 59          & 57          & 59          & 57          \\
$\mathrm{V_O}$     & 58          & 57          & 56          & 55          & 53          & 51          \\
$\mathrm{V_{Th}}$      & 56          & 55          & 53          & 52          & 50          & 47          \\
$\mathrm{O_i}$       & 57          & 56          & 54          & 52          & 52          & 50          \\
$\mathrm{Th_i}$       & 54          & 53          & 49          & 47          & 47          & 45         \\
\hline
\end{tabular*}

\end{table}

\begin{table}[htpb]
\centering
\caption{Thermal conductivity at 700 K (W/mK). }
\label{tab:conductivity}

\begin{tabular*}{\textwidth}{@{\extracolsep{\fill}}llllllll}
\hline
$x$(\%)    & \multicolumn{2}{c}{0.067} & \multicolumn{2}{c}{0.133} & \multicolumn{2}{c}{0.267} \\
\hline
Method    & RTA-NMA        & GKMA      & RTA-NMA       & GKMA     & RTA-NMA     & GKMA     \\
\hline
Pristine        & 9.32      & 8.65        & 9.32       & 8.65        & 8.65       & 8.65          \\
$\mathrm{V_O}$    & 8.74      & 8.34        & 7.93       & 7.20          & 7.01      & 6.27          \\
$\mathrm{V_{Th}}$   & 8.15      & 55       & 7.04       &6.69         & 5.93       & 5.69            \\
$\mathrm{I_O}$       & 8.47      & 7.74       & 7.72       & 7.06          & 6.86       & 6.50          \\
$\mathrm{I_{Th}}$     & 7.68    & 7.42          & 6.66       & 6.48          & 5.67       & 5.41        \\
\hline
\end{tabular*}

\end{table}

\begin{figure}[htbp]
    \centering
    \includegraphics[width=0.6\linewidth]{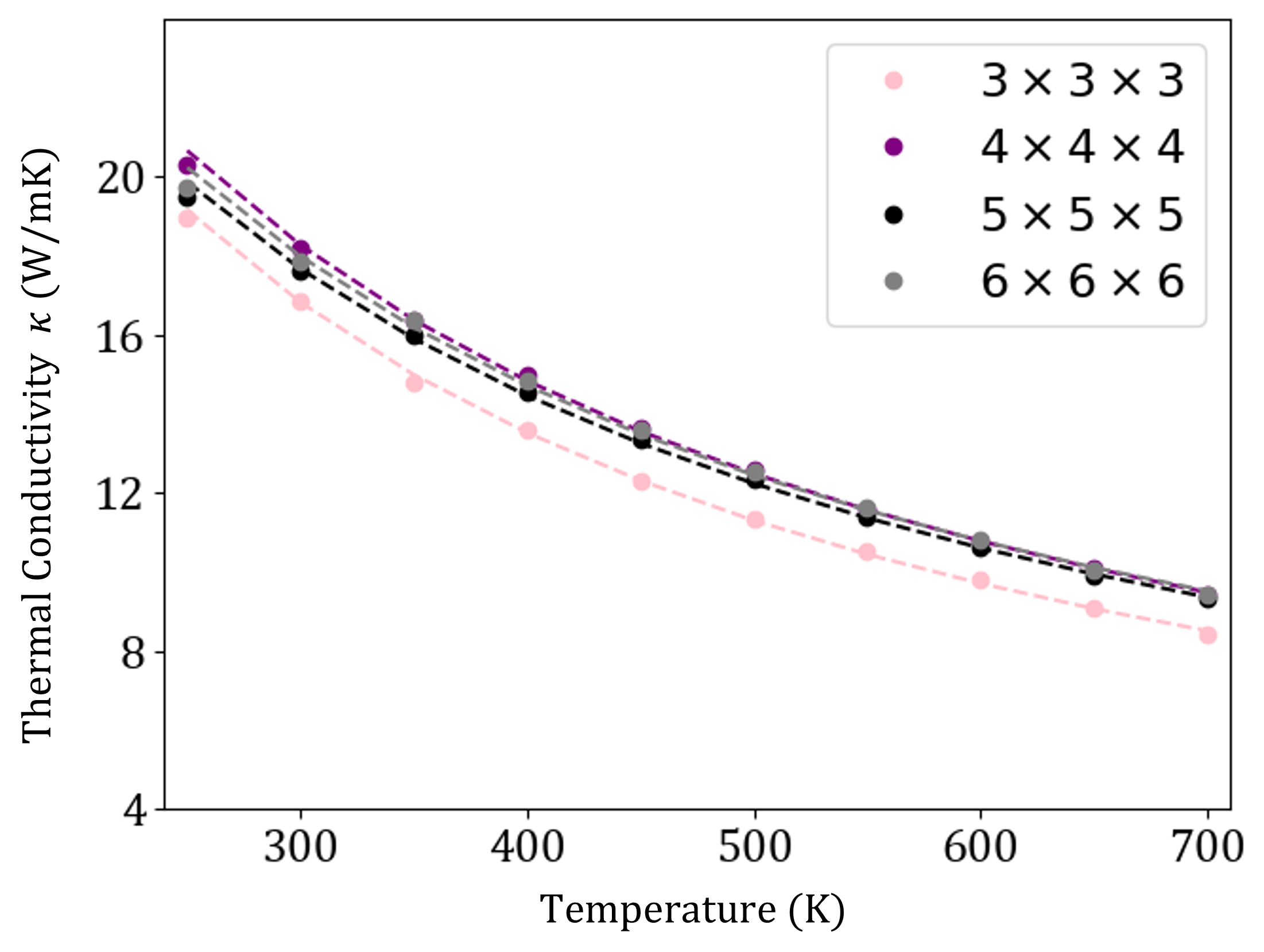}
    \caption{Thermal conductivity calculated by RTA-NMA method using supercell sizes ranging from 3$\times$3$\times$3 to 6$\times$6$\times$6 conventional unit cells, fitted $b$ terms (see Eq. (14) in the main text ) are 1.46$\times10^{-4}$, 1.27$\times10^{-4}$, 1.25$\times10^{-4}$ and 1.23$\times10^{-4}$ m/W for 3$\times$3$\times$3, 4$\times$4$\times$4, 5$\times$5$\times$5 and 6$\times$6$\times$6, respectively. }
    \label{fig:mat1}
\end{figure}

\begin{figure}[htbp]
    \centering
    \includegraphics[width=0.6\linewidth]{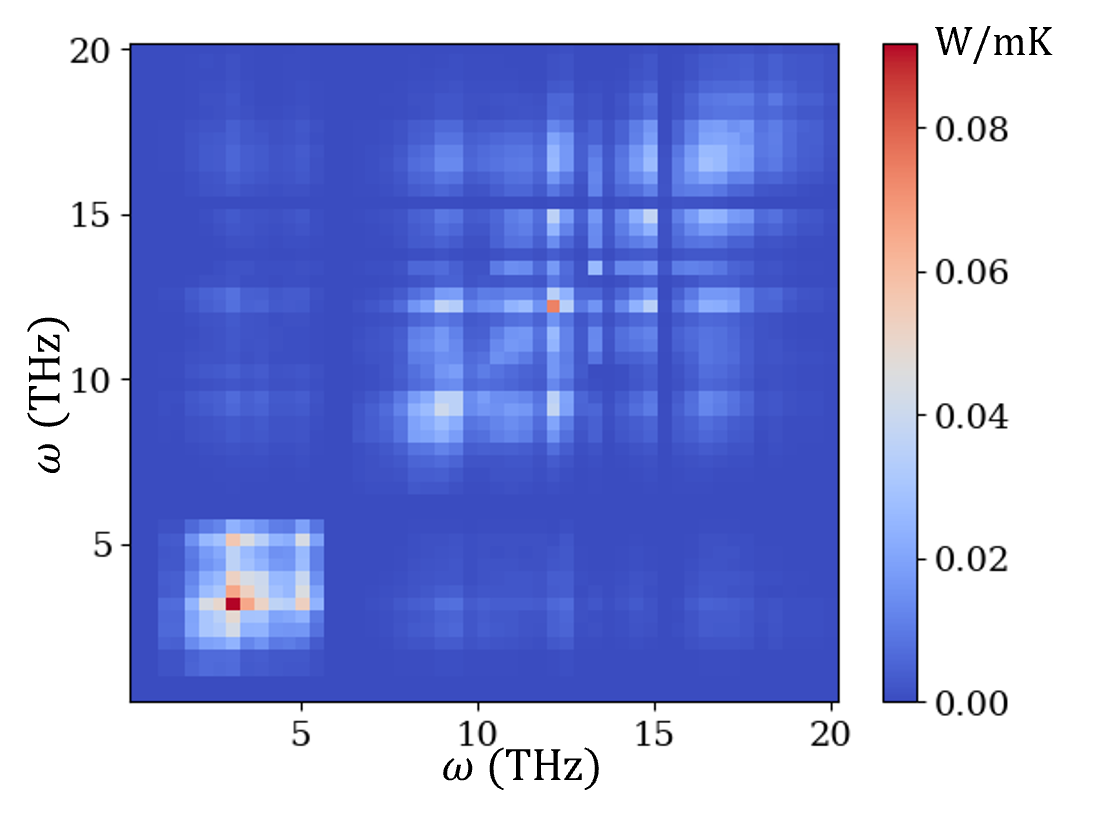}
    \caption{2D thermal conductivity map for ThO$_2$ with 0.267\% $\mathrm{V_O}$ at 300 K}
    \label{fig:mat2}
\end{figure}

\begin{figure}[htbp]
    \centering
    \includegraphics[width=0.6\linewidth]{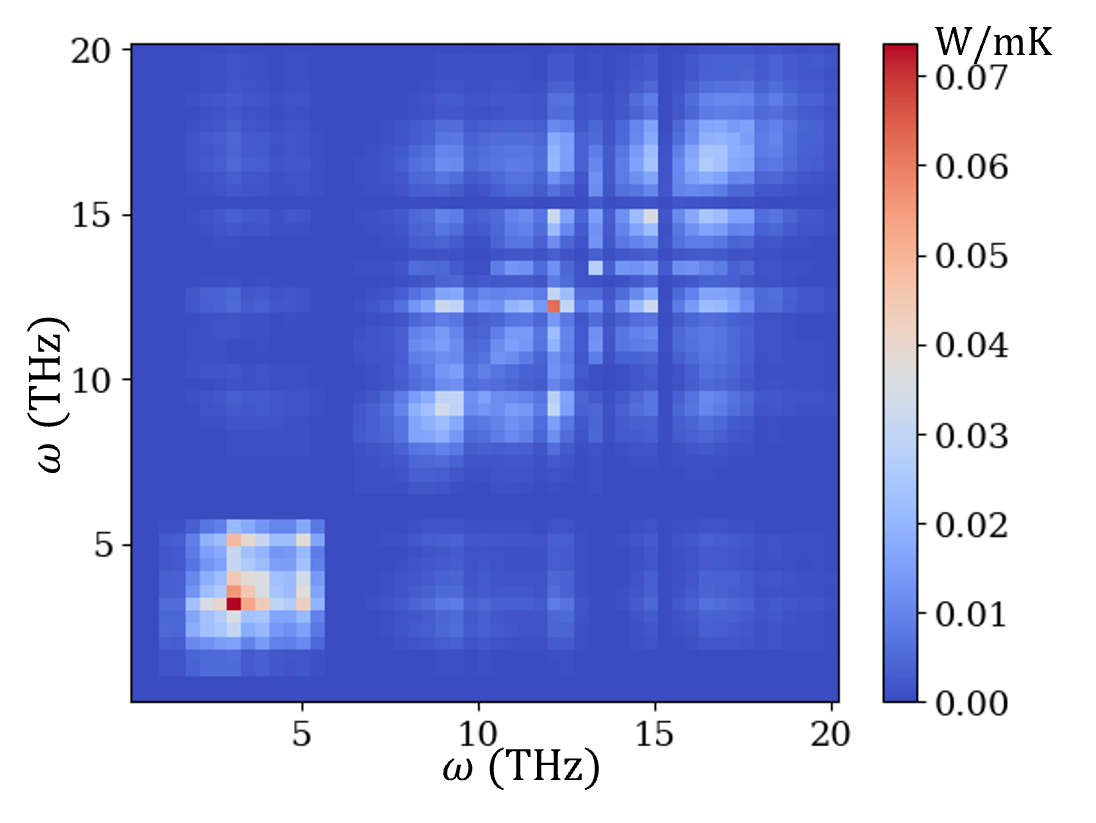}
    \caption{2D thermal conductivity map for ThO$_2$ with 0.267\% $\mathrm{V_{Th}}$ at 300 K}
    \label{fig:mat3}
\end{figure}

\begin{figure}[htbp]
    \centering
    \includegraphics[width=0.6\linewidth]{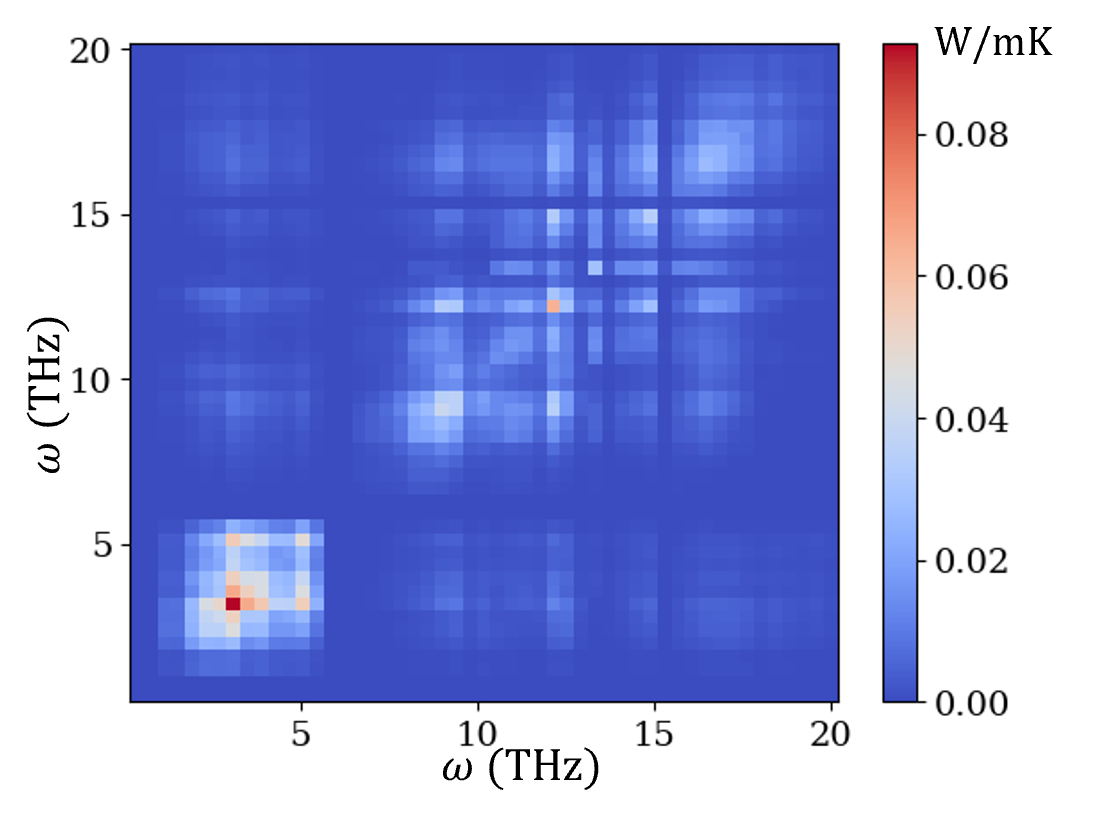}
    \caption{2D thermal conductivity map for ThO$_2$ with 0.267\% $\mathrm{I_O}$ at 300 K}
    \label{fig:mat4}
\end{figure}